\documentclass[11pt]{article}
\usepackage{setspace}
\usepackage{amssymb}
\usepackage{amsmath}
\usepackage{amsfonts}
\usepackage{slashed}

\def\be{\begin{equation}}

\def\ee{\end{equation}}

\def\dd{\partial}

\def\bea{\begin{eqnarray}}

\def\eea{\end{eqnarray}}

\newcommand\eps{\epsilon}

\makeatletter
\def\blfootnote{\xdef\@thefnmark{}\@footnotetext}
\makeatother

\begin{document}

\singlespace

\begin{flushright} BRX TH-6645 \\
CALT-TH 2019-01
\end{flushright}

\vspace*{.3in}

\begin{center}

{\Large\bf Non-Lagrangian Gauge Field Models are Physically Excluded}

{\large S.\ Deser}

{\it 
Walter Burke Institute for Theoretical Physics, \\
California Institute of Technology, Pasadena, CA 91125; \\
Physics Department,  Brandeis University, Waltham, MA 02454 \\
{\tt deser@brandeis.edu}
}
\end{center}

\begin{abstract}
While non-action-generated, but identically conserved, abelian/YM gauge vectors exist, they are unsuitable for building alternate field equations, because they have no stress-tensor, hence do not permit Poincare generators and, most physically, cannot consistently couple to gravity. Separately, their geometric analogues, covariantly conserved non-Lagrangian symmetric tensors, probably do not even exist, but their weak field, abelian, counterparts do, and share the vector fields' absence of generators.  

\end{abstract}

\section{Introduction}
This work answers an open question regarding massless gauge (abelian and non-) vector, and (less strongly) geometric tensor fields: Are there viable models with identically conserved field equations' ``left-hand-side" terms that are not derivable from actions? It is a non-trivial one, both formally and physically, as neither existence of such terms nor the proper physical grounds to exclude them are obvious; indeed it is still not known if non-singular geometrical terms even exist [1]. Vector terms do, but have no corresponding stress-tensors, hence no Poincare generators can even be defined there. More physically, they cannot consistently couple to gravity if they couple to any normal matter --- or merely if their Lagrangian counterparts are also present. Geometric tensors' (if any!) weak field versions also exist; the latter are excluded on the more formal, absence of Poincare generators, grounds. These no-go results preclude a large class of speculative models.

\section{Vectors}

A sufficiently general set of abelian vector field equations is

\begin{equation}
M^\mu = \partial_\nu \left[ X(F^2, \,^*F F) F^{\mu\nu}(A)\right] = j^\mu
\end{equation}
where $\,^*F$ is the usual dual of $F$ and the arbitrary scalar $X$ depends only on the two simplest, algebraic, invariants. 
The divergence identities $\partial_\mu M^\mu=0$ are manifest from the antisymmetry of $F^{[\mu\nu]}$ contracted with the symmetric $\partial^2_{\mu\nu}$, irrespective of $X$. However, not all such $M^\mu$ are variations of an action: they must obey the Helmholz integrability conditions, which set stringent limits on $X$. So identical conservation does NOT require an action, already in these simple examples of vectors $V^\mu=\dd_\nu H^{[\mu\nu]}$. Perhaps surprisingly, this is not a purely abelian property, but holds also for non-abelian fields: there, we replace $\partial_\mu$ by the usual covariant color derivatives $D_\mu$ whose commutator is now the non-abelian field strength, $[D_\mu, D_\nu] = F_{\mu\nu}$. Yet the generalization of (1) remains transverse, owing to the antisymmetric structure constants, since $f_{abc} F^b_{\, \, \mu\nu} F^{c \mu\nu} = 0$ (the arguments of $X$ are now the color-singlet traces of $F^2$ and $\,^*FF$). Again, only algebraic symmetry properties are relevant. Indeed, even in curved space, ordinary conservation of (1) holds, because the divergence of the contravariant tensor density $\sqrt{-g} X F^{\mu\nu}$ is still a partial derivative and so in turn is its divergence, being that of a contravariant vector density.
Are there any physically permitted models exploiting the above conservation properties, either stand-alone or by adding terms like (1) to Maxwell- or YM- like  equations? Clearly, charge conservation is not affected, since both sides of (1) are conserved. To be sure, the expression for the charge does becomes a bit byzantine, involving both the longitudinal AND transverse electric fields,
\begin{equation}
\oint d^2 {\bf S} \cdot X {\bf E} = \int d^3 x j^0 = Q.
\end{equation}

Instead, the real obstruction is due first to the loss of Poincare generators caused by the absence of an action for the $\partial (XF)$ term: no action means no conserved $T_{\mu\nu}$. For example, the divergence of a would-be $T_{\mu\nu} = X T_{\mu\nu}(\hbox{Maxwell})$ is $\sim F^2 \dd_\mu X \ne 0$; the general proof is obvious since the only possible terms are  
$A F_{\mu\alpha} F^\alpha_{\, \, \, \nu} + g_{\mu\nu} B F^2$.  Since adding non-action terms forbids stress-tensors, there are no Poincare generators; mass and spin cannot even be defined (the generators are as essential at classical as at quantum level). However, the more striking --- and physical--contradiction comes when attempting (unavoidably, if these fields are to interact with any normal ones) to couple to gravity: the added terms (while still conserved, as we saw) depend on the metric, hence are acted on but do not react on, gravity, absent a properly conserved $T_{\mu\nu}$ contribution to gravity's equations. This seeming violation of Newton's third law is not immediately inconsistent --- rather, the non-Lagrangian gauge field equation represents a sort of ``test-field": the (source-free) gravitational and gauge field equations are separate. However, if there is also a normal, say Maxwell, part --- its $T_{\mu\nu}$ is no longer conserved, and consistency is lost. Generally, if any normal matter interacts with the gauge field, its stress tensor will also no longer be conserved (on its shell) since it effectively contains the $A$-field as an ``external", rather than (normal) dynamical, parameter.
Note the contrast with Chern-Simons (CS) electrodynamics (or YM) in this respect: the CS term's stress-tensor vanishes identically, yet the original Maxwell/YM stress-tensor stays conserved on full CS shell. A large class of  speculative Maxwell and Yang-Mills extensions can thus be neglected.

\section{Gravity}
Assume the (unlikely [1]) existence of identically conserved non-Lagrangian symmetric geometric tensors 
$S_{\mu\nu}$ ($D^n$ curvature; $g_{\alpha\beta}$) and consider the physical effects of adding them to normal gravitational field equations, 
\begin{equation}
G_{\mu\nu} + S_{\mu\nu} = T_{\mu\nu}(\hbox{matter;} g),
\end{equation} 
where $G_{\mu\nu}$ denotes any Lagrangian-based tensor (or $0$) and the (normal) matter source is covariantly conserved on its shell, independent of the metric's dynamics. At linearized curvature level, where all explicit metrics as well as derivatives are flat-space, these models are similar 
to the abelian vector case: There are again identically conserved projection operators, generalizing $\partial_\nu H^{[\mu\nu]}$, namely the so-called superpotentials\footnote{In the GR literature, quantities of this type are used to represent harmless ambiguities of flat space stress tensors because they cannot contribute to any generators, whereas we use them as putative field equation contributions.
} $V^{\mu\nu}=\partial^2_{\alpha\beta} H^{[\mu\alpha] [\nu\beta]}$, where $H$ has the algebraic symmetries of the Riemann tensor. For example, in $D=2$, $H$ degenerates to $\eps^{\mu\alpha} \eps^{\nu\beta} S$, so $V_{\mu\nu}$ becomes the transverse projector $(\partial^2_{\mu\nu} -\eta_{\mu\nu} \Box) S$, where $S$ is any scalar.  Any non-Lagrangian linearized $S_{\mu\nu}$ is allowed, but as in the vector case, it has no associated stress tensor, hence loss of Poincare generators at this  linearized level --- corresponding to the Maxwell limit of the vector case. But this destroys all non-linear would-be models as well, since they would all have an abelian limit, just as YM contains Maxwell. Separately, we know [1] that any $S_{\mu\nu}$, were it to exist, starts (at least) at fourth derivative order on the curvatures, with obvious negative implications for ghost-like, and external non-Schwarzchild (if there are terms solely involving the Weyl tensor), solutions of (3).

\section{Comments}   
We have seen that while infinitely many non-Lagrangian conserved vector gauge terms exist, they are forbidden in flat space model-building owing to their obstruction to defining Poincare generators. This failure is compounded by the direct physical contradiction that they cannot consistently couple to (any) gravity, because they cannot affect the geometry as legitimate (on-shell) conserved sources, being only acted on by the metric without reacting on the latter's dynamics, not having conserved stress tensors. Yet if they are to couple to any normal matter or even if a normal, ``Maxwell", part is included, they would have to --- but cannot ---  contribute in order to insure consistency, as we have seen. Separately, while existence of conserved symmetric non-Lagrangian geometric tensors is not (yet) excluded, we noted that even if they do exist, their abelian limit encounters the corresponding vector problems.

Finally, a referee-induced comment on the use of Lagrange multipliers, the usual last resort. We could add a new vector field $B_\mu$, with a Lagrangian $L= B_\mu M^\mu$, or equivalently $L=B_{\mu\nu} X F^{\mu\nu}$, with a conserved $T_{\mu\nu}$ on $(A+B)$ shell, but it course vanishes if we set the multiplier $B=0$. The pitfall here is  that spurious degrees of freedom are introduced, as is clear when $X=1$ there $L$ describes a $2$-photon system.

\section*{Acknowledgements}
 This work was supported by the U.S. Department of Energy, Office of Science, Office of High Energy Physics, under Award Number de-sc0011632. Long-term collaboration with Y. Pang, and with A. Waldron, on a complex of related problems is happily acknowledged. I thank J. Franklin for tech help.

\end{document}